\documentclass{ws-procs975x65}

\begin{document}



\title{THE KERR THEOREM, MULTISHEETED TWISTOR SPACES
AND MULTIPARTICLE KERR-SCHILD SOLUTIONS\footnote{Talk at the GT6 session of the
MG11 meeting, supported by RFBR grant 07-08-00234 and by
travel grant from J.Sarfatti.}}

\author{ALEXANDER BURINSKII
}

\address{Gravity Research Group, NSI Russian Academy of Sciences,\\
B. Tulskaya 52, Moscow 115191, Russia, \email{bur@ibrae.ac.ru}}


\def\b{\bar}
\def\d{\partial}
\def\D{\Delta}
\def\cA{{\cal A}}
\def\cD{{\cal D}}
\def\cK{{\cal K}}
\def\cF{{\cal F}}
\def\f{\varphi}
\def\g{\gamma}
\def\G{\Gamma}
\def\l{\lambda}
\def\L{\Lambda}
\def\M{\mathcal{M}}
\def\m{\mu}
\def\n{\nu}
\def\p{\psi}
\def\q{\b q}
\def\r{\rho}
\def\t{\tau}
\def\x{\phi}
\def\X{\~\xi}
\def\~{\tilde}
\def\h{\eta}
\def\bZ{\bar Z}
\def\cY{\bar Y}
\def\bY3{\bar Y_{,3}}
\def\Y3{Y_{,3}}
\def\z{\zeta}
\def\Z{{\b\zeta}}
\def\Y{{\bar Y}}
\def\cZ{{\bar Z}}
\def\`{\dot}
\def\be{\begin{equation}}
\def\ee{\end{equation}}
\def\bea{\begin{eqnarray}}
\def\eea{\end{eqnarray}}
\def\half{\frac{1}{2}}
\def\fn{\footnote}
\def\bh{black hole \ }
\def\cL{{\cal L}}
\def\cH{{\cal H}}
\def\cP{{\cal P}}
\def\cM{{\cal M}}
\def\ol{\overline}
\def\const{{\rm const.\ }}
\def\ik{ik}
\def\mn{{\mu\nu}}
\def\a{\alpha}

\begin{abstract}
Kerr-Schild formalism is generalized by incorporation of the Kerr
Theorem with polynomials of higher degrees in $Y\in CP^1.$ It
leads to multisheeted twistor spaces and multiparticle solutions.
\end{abstract}

\bodymatter

{\bf 1.}{\bf Introduction.} The Kerr-Newman solution displays many
relationships to the quantum world. It is the anomalous
gyromagnetic ratio $g=2$,  stringy structures  and other features
allowing one to construct a semiclassical model of the extended
electron \cite{Bur0,BurAxi,BurTwi} which has the Compton size and
possesses the wave properties.

One of the mysteries of the Kerr geometry is  the existence of two
sheets of space-time, $(+)$ and $(-)$, on which the dissimilar
gravitation (and electromagnetic) fields are realized, and fields
living on the $(+)$-sheet do not feel the fields of the
$(-)$-sheet. Origin of this twofoldedness lies in the Kerr
theorem, generating function $F$ of which for the Kerr-Newman
solution has two roots which determine two different twistorial
structures on {\it the same} space-time.

The standard Kerr-Schild formalism is based on a restricted
version of the Kerr theorem which uses polynomials of second
degree in $Y,$ and, in fact, produced only the Kerr geometry.
The use of Kerr theorem in full power is related with the
treatments of polynomials of higher degrees in $Y.$ On this way we
obtain the multisheeted twistor spaces and corresponding
multiparticle Kerr-Schild solutions \cite{wonder,Multiks}.
The case of a quadratic in $Y$ generating function of the Kerr
Theorem $F(Y)$ was investigated in details in
\cite{BurMag,BurNst}. It leads to the Kerr spinning
particle (or black hole) with an arbitrary position, orientation
and boost. Choosing generating function $F(Y)$  as a product of
partial functions $F_i$ for spinning particles i=1,...k, we obtain
multi-sheeted, multi-twistorial space-time over $M^4$ possessing
unusual properties. Twistorial structures of the i-th and j-th
particles turn out to be independent, forming a type of its
internal space. However, the exact solutions show that gravitation
and electromagnetic interaction of the particles occurs via the
connecting them singular twistor lines. The space-time of the
multiparticle solutions turns out to be covered by a net of
twistor lines, and  we conjecture that it reflects its relation to
quantum gravity.

Recall that {\bf the Kerr-Newman metric} can be represented in the
Kerr-Schild form $g_{\m\n} = \h_{\m\n} + 2 h k_{\m} k_{\n},$
where $ \h_\mn $ is metric of auxiliary Minkowski space-time, and
$ h=(mr -e^2/2)/(r^2+a^2\cos ^2 \theta).$
 $k_\m (x)$ is a twisting null field, which is tangent to the
Kerr principal null congruence (PNC) which is geodesic and
shear-free \cite{DKS,KraSte,BurNst}. PNC is determined by the complex
function $Y(x)$ via the one-form \be
 e^3 = du+ \Y d \z  + Y d \Z - Y \Y d v  = P k_\m dx^\m
\label{cong} \ee where $u, \ v, \ \Z , \ \z $ are the null
Cartesian coordinates. Here $P$ is a normalizing factor for $k_\m$
which provide $k_0 =1$ in the rest frame. The null rays of the Kerr congruence
are twistors.

{\bf The Kerr theorem} \cite{DKS,KraSte} allows one to describe the
Kerr geometry in twistor terms \cite{BurNst}.

It claims that any geodesic and shear-free null congruence in
Minkowski space-time is defined by a function $Y(x)$ which is a
solution of the equation \be F  = 0 , \label{KT}\ee where $F
(Y,\l_1,\l_2)$ is an arbitrary holomorphic function of the
projective twistor coordinates \be Y,\quad \l_1 = \z - Y v, \quad
\l_2 =u + Y \Z .\label{Tw}  \ee

In the Kerr-Schild backgrounds the Kerr theorem acquires a more
broad content, allowing one  to determine the normalizing function
$P$ and complex radial distance $\tilde r=r+ia\cos\theta ,$

$ P = \d_{\l_1} F - \Y \d_{\l_2} F , \quad \tilde r=PZ^{-1}= -
\quad d F / d Y  $,and therefore, restore all
the necessary characteristics of the corresponding solutions,
including the electromagnetic
field of the corresponding Einstein-Maxwell equations up to an
arbitrary function. The
position of singular lines, caustics of PNC, corresponds to
$\tilde r=0$, and is determined by the system of equations $
F=0;\quad d F / d Y =0 \ .$

{\bf Multi-twistorial space-time.} Selecting an isolated i-th
particle with parameters $q_i$, one can obtain the roots $Y_i^\pm
(x)$ of the equation $F_i(Y|q_i)=0$ and express $F_i$ in the form
$ F_i(Y)=A_i(x)(Y-Y_i^+)(Y-Y_i^-).$ Then,
substituting the $(+)$ or $(-)$ roots $Y_i^\pm (x)$ in the
relation (\ref{cong}), one determines congruence
$k^{(i)}_{\mu}(x)$ and consequently, the Kerr-Schild ansatz
for metric
$ g^{(i)}_\mn =\eta _\mn + 2h^{(i)} k^{(i)}_{\mu} k^{(i)}_{\nu} ,$

and finally, the function $h^{(i)}(x)$ may be expressed in terms
of $\tilde r_i= - d_Y F_i .$

What happens if we have a system of $k$ particles? One can form
the function $F$ as a product of the known blocks $F_i(Y)$,
$ F(Y)\equiv \ \prod _{i=1}^k F_i (Y) \label{multi}. $
The solution of the equation $F=0$ acquires $2k$ roots $Y_i^\pm$,
and the twistorial space turns out to be multi-sheeted.
The twistorial structure on the i-th $(+)$ or $(-)$ sheet is
determined by the equation $F_i=0$ and does not depend on the
other functions $F_j , \quad j\ne i$. Therefore, the particle $i$
does not feel the twistorial structures of other particles.
Similar, the condition for singular lines $F=0, \ d_Y F=0$
acquires the form

\be \prod _{l=1 }^k F_l =0, \qquad   \sum ^k_{i=1} \prod _{l\ne
i}^k F_l d_Y F_i =0 \label{leib} \ee
 and splits into k independent relations
$ F_i=0,\quad \prod _{l\ne i}^k F_l d_Y F_i =0 .$

One sees, that i-th particle does not feel also singular lines of
other particles.  The  space-time splits on the independent
twistorial sheets, and therefore, the twistorial structure related
to the i-th particle plays the role of  its ``internal space''.
It looks wonderful. However, it is a direct generalization of the
well known twofoldedness of the Kerr space-time which remains one
of the mysteries of the Kerr solution for the very long time.
 The negative sheet of Kerr geometry may be treated as the sheet
of advanced fields. In this case the source of spinning particle
turns out to be the Kerr singular ring (circular string,
\cite{BurAxi,BurTwi}) with the electromagnetic excitations in the
form of traveling waves which generate spin and mass of the
particle (microgeon model  \cite{Bur0,BurTwi}).

{\bf Multi-particle Kerr-Schild solution.} Using the Kerr-Schild
formalism with the considered above generating functions $\prod
_{i=1}^k F_i (Y)=0,$ one can obtain the exact asymptotically flat
multi-particle solutions of the Einstein-Maxwell field equations.
Since congruences are independent on the different sheets, the
congruence on the  i-th sheet retains to be geodesic and
shear-free, and one can use the standard Kerr-Schild algorithm of
the paper \cite{DKS}. One could expect that result for the i-th
sheet will be in this case the same as the known solution for
isolated particle. Unexpectedly, there appears a new feature
having a very important consequence.

In addition to the usual Kerr-Newman solution for an isolated
spinning particle, there appears a series of the exact `dressed'
Kerr-Newman solutions which take into account surrounding
particles and differ by the appearance of singular twistor strings
connecting the selected particle to external particles. This is a
new gravitational phenomena which points out on a probable stringy
(twistorial) texture of vacuum and may open a geometrical way to
quantum gravity.

\vfill

\end{document}